\documentclass{article}
\usepackage{spconf,amsmath,graphicx}
\usepackage{cite}
\usepackage{epstopdf}
\usepackage{amssymb}
\usepackage{caption}
\usepackage{tabularx}
\usepackage{booktabs}
\usepackage{algorithm}
\usepackage{algorithmic}
\usepackage{multirow}
\usepackage{stfloats}
\usepackage{setspace}
\usepackage{booktabs}
\usepackage{graphicx}
\usepackage{subcaption}

\title{PDPCRN: Parallel Dual-Path CRN with Bi-directional Inter-Branch Interactions for Multi-Channel Speech Enhancement}

\name{Jiahui Pan$^{*}$, Shulin He$^{*}$\thanks{*: Equal Contribution.}, TianciWu, Hui Zhang$^{\dag }$\thanks{\dag: Corresponding author.}, Xueliang Zhang}

\address{
College of Computer Science, Inner Mongolia University, China\\
\texttt{panjiahui@mail.imu.edu.cn, \{cszh,cszxl\}@imu.edu.cn}
}

\begin{document}
\setstretch{0.96}
\small
\maketitle
\begin{abstract}
Multi-channel speech enhancement seeks to utilize spatial information to distinguish target speech from interfering signals. While deep learning approaches like the dual-path convolutional recurrent network (DPCRN) have made strides, challenges persist in effectively modeling inter-channel correlations and amalgamating multi-level information. In response, we introduce the Parallel Dual-Path Convolutional Recurrent Network (PDPCRN). This acoustic modeling architecture has two key innovations. First, a parallel design with separate branches extracts complementary features. Second, bi-directional modules enable cross-branch communication. Together, these facilitate diverse representation fusion and enhanced modeling. Experimental validation on TIMIT datasets underscores the prowess of PDPCRN. Notably, against baseline models like the standard DPCRN, PDPCRN not only outperforms in PESQ and STOI metrics but also boasts a leaner computational footprint with reduced parameters.

\end{abstract}
\section{Introduction}
Multi-channel speech enhancement aims to extract a clean speech signal from background noise using multiple microphone recordings.
Effectively exploiting the spatial and inter-channel information is critical for enhancing clean speech. 
Multi-channel signal provides spatial diversity \cite{hawley2004benefit}, enabling the capture of useful spatial cues encoded in the phase and amplitude relationships between microphones. This additional spatial information can lead to significant performance improvements over single-channel approaches. 
However, efficiently modeling the inter-channel correlations and integrating multi-level contextual information continue to be significant hurdles in the field.

Traditional approaches for multi-microphone speech enhancement utilize spatial filtering methods \cite{benesty2008microphone,gannot2017consolidated} that leverage spatial information from the acoustic environment, including the angular position of the target speech and microphone array geometry. 
These methods, commonly termed beamforming, these techniques apply linear processing to weight the individual microphone channels in the time-frequency domain, with the goal of suppressing signal components that do not originate from the desired source. 
Classic beamforming algorithms such as the delay-and-sum beamformer \cite{klemm2008improved}, minimum variance distortionless response (MVDR) \cite{souden2009optimal} beamformer, and super-directivity beamformer \cite{bitzer2001superdirective} can yield strong performance. 
However, they rely heavily on accurate estimation of spatial information, which remains a challenging task under noisy conditions.

With the emergence of deep learning, various deep neural network (DNN) architectures have been developed for multi-channel speech enhancement. 
Typical neural networks, including convolutional neural networks (CNNs) \cite{li2021two}, recurrent neural networks (RNNs) \cite{sun2017multiple}, and more recently attention mechanisms \cite{kim2020t}, have been successfully applied to time-frequency \cite{xu2014regression,wang2018supervised} and time-domain \cite{pascual2017segan,fu2018end} speech enhancement. 
Leveraging both CNN and RNN strengths, the proposed convolutional recurrent network (CRN) \cite{zhao2018convolutional} with its convolutional encoder-decoder structure and recurrent bottleneck has become popular for real-time speech enhancement \cite{2020Learning}. To address long sequence modeling challenges, the dual-path recurrent neural network (DPRNN) \cite{molchanov2016pruning} was proposed, where long sequential features are divided into smaller chunks and recursively processed by intra-chunk and inter-chunk RNNs, thereby reducing the sequence length handled per RNN.
While the dual-path convolutional recurrent neural network (DPCRN) proposed by Le et al. \cite{le2021dpcrn} aims to integrate the strengths of CNNs for local pattern modeling and DPRNNs for long-term modeling, it relies solely on spectral input without explicit spatial features.

The spatial information is vital for multi-channel scenarios. However, without explicit spatial features, the above approaches only use multi-channel spectra as input, relying on implicit learning of spatial information.
To address this gap, we present PDPCRN to effectively model inter-channel correlations and incorporate multi-level information via two key innovations.
\textbf{Parallel Structure with Dual Branches:} This design seeks to harness complementary features from the input.
The first branch (DPRNN + Self-Attention): The self-attention mechanism \cite{shaw2018self} is pivotal in recognizing and weighting critical portions of the input, facilitating the model to prioritize salient features and downplay less pertinent ones. By integrating self-attention with the DPRNN's known capabilities in modeling long-term temporal dependencies, this branch specializes in highlighting the most relevant speech components.
The second branch (Depthwise Convolutions + DPRNN): Depthwise convolutions \cite{han2021demystifying} are pivotal for feature extraction. Unlike standard convolutions, depthwise convolutions operate on individual channels, making them adept at capturing spatial localization cues from multi-channel data. Importantly, they achieve this without substantially incrementing model parameters, ensuring computational efficiency.
\textbf{Bi-directional Interaction Module:} This component is the cornerstone for mutual learning. By allowing branch outputs to be reciprocally passed, the two branches inform and refine each other's feature representations. 
This approach addresses the limitations of existing architectures that lack interaction between channel and spatial information. 
It ensures that the inter-channel correlations and multi-level information are seamlessly integrated, with the branches complementing each other's strengths.
Evaluations on TIMIT under varying noise and reverberation show our model outperforms established benchmarks. Remarkably, this is achieved with fewer computations and parameters. By modeling of inter-channel correlations and integration of multi-level information, our model efficiently improves the performance of the multi-channel speech enhancement.

\section{PROPOSED METHODS}
\begin{figure*}[ht]
	\centering
	\includegraphics[width=1.0\textwidth]{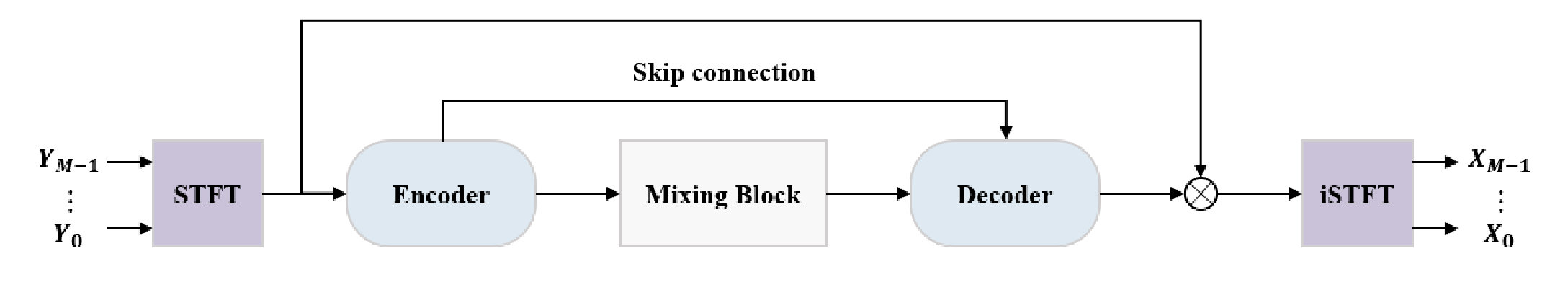}
	\caption{The architecture of the proposed PDPCRN system.}
	\label{fig:fig1}
\end{figure*}

\begin{figure}
    \centering
    \includegraphics[width=0.5\textwidth]{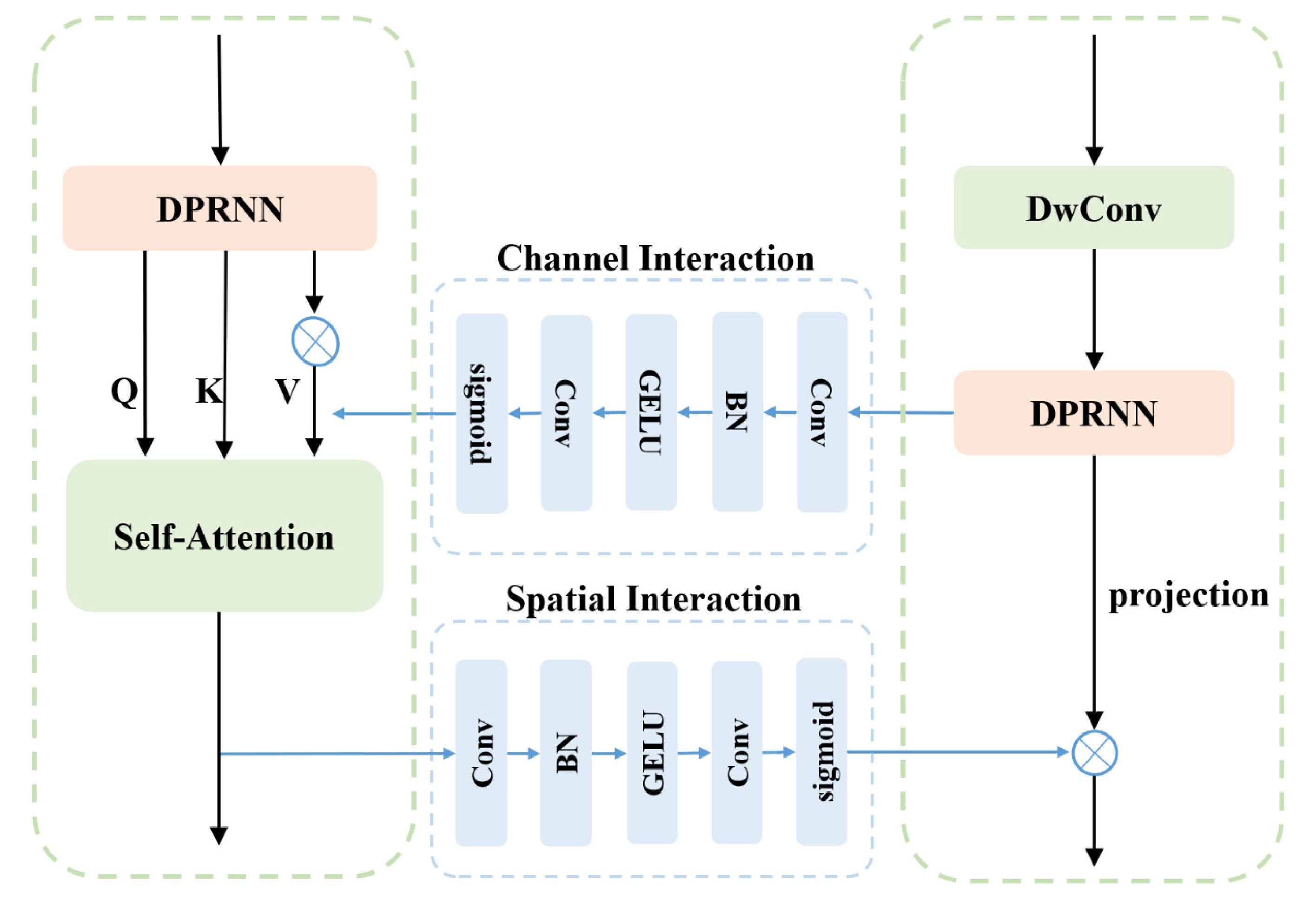}
    \caption{The detail of Mixing Block.}
    \label{fig:fig2}
\end{figure}

The proposed PDPCRN architecture, as illustrated in Fig.\ref{fig:fig1} and Fig.\ref{fig:fig2}, comprises two primary novel components: (1) A parallel structure with distinct DPRNN branches complemented by self-attention and depthwise convolution to extract hierarchical features. (2) Bi-directional connections between the branches to enable cross-branch feature sharing and representation enhancement in both pathways. The dual-branch design with tailored modules extracts multi-level representations, while the inter-branch interactions further enrich the learned features in each branch. Together, these innovations in the proposed PDPCRN model facilitate advanced inter-channel correlation modeling and integration of multi-level information for speech enhancement.

We adopt a Multiple Input Multiple Output (MIMO) architecture as illustrated in Fig.\ref{fig:fig1}. The input comprises mixed speech signals from M microphone channels. The model predicts M target speech outputs, one corresponding to each of the M input channels. We consider an array with M microphones. The sound captured at the $m$-th microphone signal can be decomposed as:
\begin{equation}
	y_{m}(n)= h_{m}(n)*x_{m}(n)+v_{m}(n),
\end{equation}
where $x_{m}(n)$ denotes the direct speech component in the $m$-th microphone signal corresponding to speech, $v_{m}(n)$ denotes the noise component representing reverberation, background noise and any remaining components and $n$ denotes the discrete time index.
By N-point short-time Fourier transform(STFT), the $y_{m}(n)$ in the T-F domain can be recorded as $Y_{m}(t, f)$:
\begin{equation}
Y_{m}(t, f)=H_{m}(t, f) X_{m}(t, f)+V_{m}(t, f),
\end{equation}
where $t$ and $f$ are the frame index and the frequency index.
$X_{m}(t, f)$ is the STFT of $x_{m}(n)$, $V_{m}(t, f)$ denotes the noise component. 
Considering the symmetry of $Y_{m}(t, f)$ in frequency.

\subsection{The Parallel Design.}
In this work, we propose a parallel design for hierarchical feature learning. As depicted in Fig.\ref{fig:fig2}, this parallel design enables interweaving of features across branches for enhanced representation learning, with self-attention and depthwise convolutions in separate pathways.
The first branch employs a DPRNN module combined with a self-attention mechanism. 
DPRNN divides the long input sequence into smaller chunks, with an intra-chunk RNN and inter-chunk RNN applied recursively to model local and global dependencies. The self-attention mechanism highlights salient parts of the input to focus on extracting speech components.
The second branch utilizes depthwise convolutions followed by DPRNN. 
Depthwise convolutions efficiently learn spatial localization cues from the multi-channel input without substantially increasing parameters. The lightweight convolutions extract inter-channel features before DPRNN models temporal dependencies.

\textbf{Enhanced Hierarchical Feature Representation.}
The presented parallel architecture presents advantages concerning hierarchical feature representation. The branching structure synergizes self-attention and depthwise convolutions to intricately connect spatial and temporal dependencies across different tiers.
Specifically, 
self-attention models longer-range dependencies and global context at higher layers of the hierarchy.
Conversely, 
depthwise convolutions provide more localized spatial patterns at lower layers
The interleaving of these dual pathways amalgamates localized features with global context, culminating in a more intricate hierarchical representation. 
This approach facilitates the encapsulation of meticulous spatial cues alongside comprehensive temporal significance, operating at diverse levels of abstraction.

\textbf{Computational Efficiency.}
The DPRNN modules partition the input sequence into segments, mitigating computational demands associated with modeling extensive sequences.
This enhancement bolsters efficiency during the handling of prolonged inputs. Furthermore, the utilization of lightweight depthwise convolutions curbs computation by circumventing excessive parameter expansion. In synergy, these pathways strike a harmonious equilibrium between computational expense and representational prowess. DPRNN efficiently manages memory and computation for temporal modeling, while streamlined spatial convolutions extract localized patterns without substantially increasing model size. The collaborative architecture allows comprehensive exploration of spatial and temporal interdependencies within a computationally efficient framework, enabled by the hierarchical parallel design.

\subsection{Bi-directional Interactions.}
We introduce a bi-directional interactions module that facilitates the exchange of outputs between the two branches. This interaction enables mutual learning and reinforcement between branches, augmenting their representations. This process is visually depicted in Fig.\ref{fig:fig2}. Notably, the channel interaction mechanism transmits information from the right branch to the left one, thereby amplifying channel modeling. Simultaneously, spatial interactions propagate spatial relationships from the left branch to the right one. This integration injects speech context, thereby assisting in the inter-channel feature learning of the second branch.

Within the bi-directional interaction module, both channel and spatial components are incorporated. The channel module encompasses a pair of $2\times 2$ convolutional layers, succeeded by batch normalization (BN) \cite{ioffe2015batch} and GELU \cite{hendrycks2016gaussian} activation. This configuration yields channel attention maps, which in turn facilitate the dissemination of information, thereby enriching channel modeling. The spatial module adheres to the same architectural design.
It is important to highlight two key aspects:
\begin{itemize}
    \item The flow of information from both the DPRNN and self-attention branches is directed towards the right branch through spatial interaction, a process that occurs subsequent to the module's output.
    \item Information originating from the deep convolution and DPRNN branches is directed to the self-attention value of the left branch through channel interaction.
\end{itemize}

\section{EXPERIMENTS}
\subsection{Datasets and Evaluation Metrics}
\label{ssec:Dataset}

The training data are synthesized by convolving multi-channel room impulse responses (RIRs) \cite{allen1979image} with diverse speech signals extracted from the TIMIT \cite{zue1990speech} dataset. The clean segments in the TIMIT database are categorized into three exclusive subsets: training, validation, and testing. Noise segments from the DNS-Challenge corpus are employed for both training and validation, whereas the testing set utilizes NOISEX-92 \cite{varga1993assessment} and cafe noises from CHiME3 \cite{barker2015third}. 
In the data generation phase, relying on a uniform circular array comprising 16 omnidirectional microphones. The array radius is 0.035 m, with a random placement inside the room, while maintaining a source-to-array center distance of 1 m. The generated RIRs pertain to a room with dimensions of $6 \times 5 \times 4 \mathrm{~m}^{3}$, characterized by a variety of SNR and reverberation time RT60 values. SNR values range from -10 dB to 10 dB, while RT60 values span from 0.2 seconds to 1.0 second. Overall, around 24,000 and 2,600 multichannel reverberant noisy mixtures are generated for training and validation, respectively. 
For the purpose of evaluation, we define five distinct SNR levels: -10dB, -5dB, 0dB, 5dB, and 10dB. In addition, we explore nine distinct T60 values, ranging from 0.2s to 1.0s, with intervals of 0.1s. This comprehensive configuration results in the generation of 350 pairs for each specific case.

The study used two primary metrics to evaluate model performance: perceptual evaluation of speech quality (PESQ) \cite{rix2001perceptual} and short-time objective intelligibility (STOI) \cite{taal2011algorithm}. PESQ rates speech quality on a scale from -0.5 to 4.5, while STOI gauges speech intelligibility on a scale of 0 to 100. Improved scores in both metrics reflect better performance.

\subsection{Experiment Setup}
In our model, the encoder is composed of convolutional layers with channel configurations:\{32, 32, 32, 64, 80\}. 
The kernel size and the stride are respectively set to \{(2,5),(2,3),(2,3),(2,3),(2,3)\} and \{(1,2),(1,2),(1,1),(1,1),(1,1)\} in frequency and time dimension. 
Causal computation is utilized across all Conv-2D and transposed Conv-2D layers. 
The architecture involves the utilization of two Mixing Blocks, with each block encompassing two DPRNNs, a multi-head self-attention mechanism, and a depthwise convolution. Specifically, the self-attention mechanism employs 50 heads, and the depthwise convolution is characterized by a kernel size of $1 \times 3$. To ensure comparable computational complexity and parameter volume, each DPRNN module is equipped with RNNs featuring a hidden dimension of 80. Likewise, the input feature dimension of the depthwise convolution is set to 80.

All the utterances are sampled at 16 kHz, we have configured the window length as 25 ms and the hop size as 12.5 ms. An FFT length of 400 is employed, with the application of a sine window prior to the execution of FFT and overlap-add operations. The input to the model comprises a 201-dimensional complex spectrum. Adam optimizer is applied with the initial learning rate set to 1e-3. If validation loss does not decrease for consecutive two epochs, the learning rate will be halved. All models are trained for 60 epochs.

\begin{table}[t]
\centering
  \caption{Comparisoins if different approaches in Params and FLOPs.}
    \begin{tabular}{@{}ccc@{}}
        \toprule
        Method & \#Params(K) & FLOPs(G) \\ \midrule
        DPCRN & 814.60 & 3.09 \\
        PDPCRN & \textbf{790.78} & \textbf{3.05} \\ \bottomrule
    \end{tabular}
    \label{Params}
\end{table}

\subsection{Results and analysis}
In this study, the proposed PDPCRN is compared with GCRN \cite{tan2019learning} and DPCRN architectures. The GCRN employs convolutional recurrent networks for complex spectral mapping. It is designed to map from real and imaginary spectrograms of noisy speech to their clean counterparts, consequently enhancing both magnitude and phase responses of speech. On the other hand, DPCRN is a speech enhancement model in the time-frequency domain. This model combines the local pattern modeling capability of CNNs with the long-term sequence modeling capacity of DPRNNs.

\subsubsection{Performance for Proposed System}
Table \ref{label_summary} delineates the performance across varying SNRs. Our PDPCRN model showcases enhancements over both baselines across different signal-to-noise ratios. Specifically, in comparison to GCRN, there is a 7.5\% relative improvement in the PESQ average column and a 4.6\% advancement in the STOI average column. When contrasted with DPCRN, the relative gains are 3.9\% for the PESQ average and 3.4\% for the STOI average.
Furthermore, Table \ref{Params} presents a comparison between the computational load and parameter count of the proposed model and DPCRN. Notably, while our model requires fewer computations and has a smaller parameter count than DPCRN, it still delivers superior performance.
\begin{table*}[t]
\centering
  \caption{Comparisoins of different approaches in PESQ and STOI.}
    \begin{tabular}{ccccccccccccc}
   \toprule
   \multirow{2}{*}[-2pt]{\textbf{Methods}} &
   \multicolumn{6}{c}{\textbf{PESQ}} & \multicolumn{6}{c}{\textbf{STOI (in \%)}} \\
   \cmidrule(r){2-7} 
   \cmidrule(l){8-13}
   & -10 dB & -5 dB & 0 dB  & 5 dB & 10 dB & Avg. & -10 dB & -5 dB & 0 dB & 5 dB  & 10 dB & Avg.\\
   \midrule
Unprocessed
     &1.25 &1.30 &1.42 &1.62 &1.85 &1.49 &39.14 &47.64 &56.95 &66.39 &75.02 &57.03\\
GCRN
     &1.33 &1.50 &1.68 &1.97 &2.21 &1.74 &43.72 &56.34 &66.57 &75.57 &82.04 &64.85\\
DPCRN
     &1.34 &1.54 &1.76 &2.07 &2.29 &1.80 &43.19 &56.39 &67.72 &77.04 &83.71 &65.61\\
\midrule

\textbf{PDPCRN}		
      & \textbf{1.36} & \textbf{1.57} & \textbf{1.84} & \textbf{2.17} & \textbf{2.41} & \textbf{1.87}    
      & \textbf{45.20} & \textbf{59.00} & \textbf{70.33} & \textbf{79.23} &  \textbf{85.32} & \textbf{67.82} \\
\bottomrule
  \end{tabular}
  \label{label_summary}
\end{table*}

\begin{figure}
    \centering
    \includegraphics[width=0.4\textwidth]{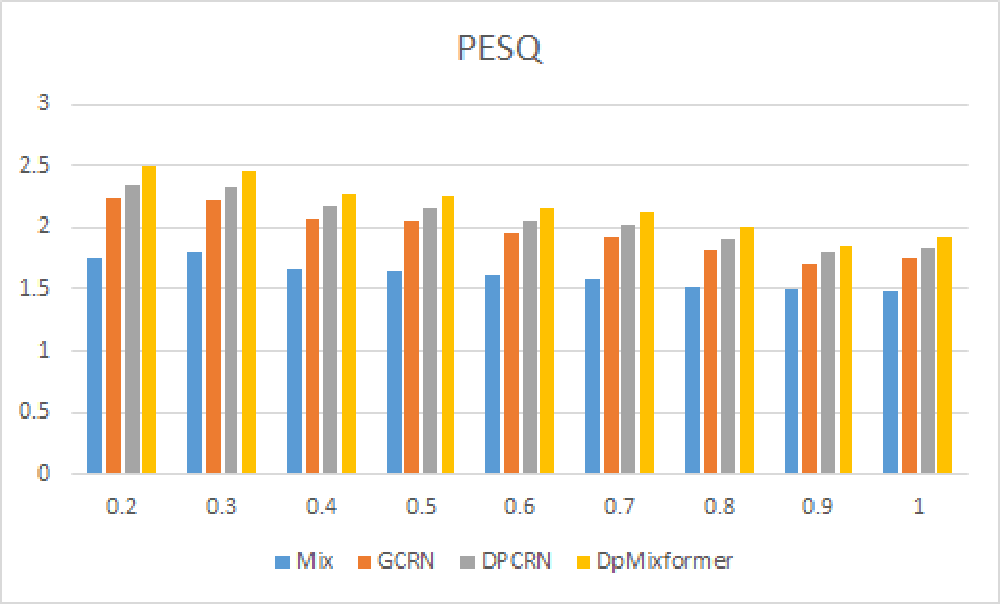}
    \caption{PESQ Comparison of Models at -5dB SNR Across Different RT60 Values.}
    \label{fig:PESQ-5}
\end{figure}

\begin{figure}
    \centering
    \includegraphics[width=0.4\textwidth]{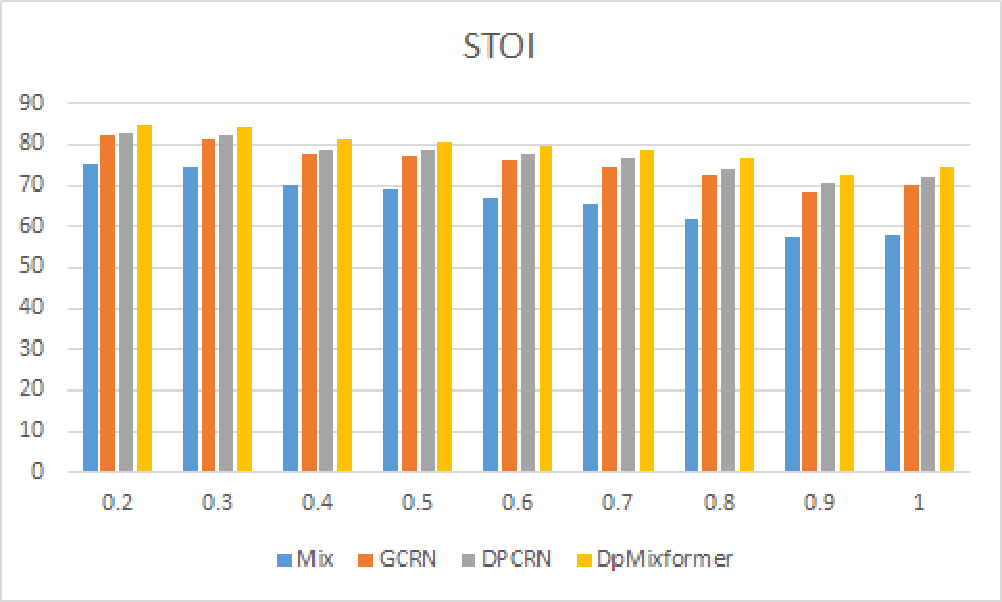}
    \caption{STOI Comparison of Models at -5dB SNR Across Different RT60 Values.}
    \label{fig:STOI}
\end{figure}

We extended our comparison to assess the performance of the proposed model under various reverberation conditions. As illustrated in Fig.\ref{fig:PESQ-5} and Fig.\ref{fig:STOI}, evaluations were conducted at SNR levels of -5dB, with RT60 ranging from 0.2s to 1.0s. The findings indicate that our method surpasses the baseline approach, particularly excelling in scenarios with lower SNR and higher reverberation. Specifically, at RT60 of 1.0s, our model yielded a 5.8\% relative improvement in STOI when compared to DPCRN.

\subsubsection{Ablation Study}
We further evaluate the influence of the bi-directional interactions module through ablation experiments. As shown in Table \ref{tab:Ablation}, PDPCRN without bi-directional interactions (PDPCRN w/o BI) exhibits degraded performance. Experiments were conducted with RT60 of 0.2s and SNR levels of [-10, -5, 0, 5, 10] dB. Notably, PESQ scores decrease at -10, 0, and 10 dB without bi-directional interactions, with the largest drop from 1.41 to 1.39 at -10 dB. Similarly, STOI results confirm the importance of bi-directional interactions, with performance declining in its absence. When evaluated at an SNR of -10 dB, STOI drops from 50.62\% to 49.88\% without the module. Overall, the results demonstrate the significance of bi-directional interactions for representation learning.

\begin{table}[h]
\Huge
\centering
  \caption{Comparisons of different approaches in STOI and PESQ at 0.2s RT60.}
 
  \label{tab:Ablation}
  \resizebox{\linewidth}{!}{
    \begin{tabular}{ccccccc}
    \toprule
    \multirow{2}{*}[-2pt]{\textbf{Methods}} &
    \multicolumn{2}{c}{\textbf{PESQ}} & \multicolumn{2}{c}{\textbf{STOI (in \%)}} \\
    \cmidrule(r){2-3} 
    \cmidrule(l){4-5}
     & PDPCRN (w/o BI) & PDPCRN  & PDPCRN (w/o BI) & PDPCRN\\
    \midrule
    -10dB  & 1.39 & \textbf{1.41}  & 49.88 & \textbf{50.62} \\
    -5dB & 1.67 & 1.67  & 64.05 & \textbf{64.30} \\
    0dB  & 2.02 & \textbf{2.03}  & 75.50 & \textbf{76.12} \\
    5dB  & 2.49 & 2.49 & 82.69  & \textbf{84.77} \\
    10dB  & 2.86 & \textbf{2.87}  & 90.47 & \textbf{90.80} \\
    \bottomrule
    \end{tabular}
    }   
\end{table}

\section{CONCLUSION}
In this paper, we propose a Parallel Dual-Path Convolutional Recurrent Network (PDPCRN) for multi-channel speech enhancement. The key novelty lies in the parallel dual-path architecture integrated with bi-directional interaction modules. This enables efficient extraction of spatial information and fusion of multi-level information. Experiments demonstrate the PDPCRN effectively enhances both spectral and spatial attributes of speech, highlighting the importance of joint spectral-spatial optimization. Moreover, analysis of the bi-directional interaction module reveals its critical role in facilitating efficient cross-branch information fusion, leading to improved model performance. In summary, the parallel dual-path structure with bi-directional interactions allows combined spectral-spatial optimization for effective multi-channel speech enhancement.

\section{Acknowledgements}
This work was partly supported by the China National Nature Science Foundation (No. 61876214, No. 61866030).

\newpage
\small
\setstretch{0.9}
\bibliographystyle{IEEEbib}
\bibliography{strings,refs}

\end{document}